# Edge effect and significant increase of the superconducting transition onset temperature of 2D superconductors in flat and curved geometries

Chi Ho Wong[1,2] and Rolf Lortz[1*]

[1]*Department of Physics, The Hong Kong University of Science and Technology, Clear Water Bay, Kowloon, Hong Kong*

[2]*Institute of Physics and Technology, Ural Federal University, Russia*

[*]*Corresponding author: lortz@ust.hk*

## 1 Abstract

In this paper, we present a simple method to model the curvature activated phonon softening in a 2D superconducting layer. The superconducting transition temperature $T_c$ in the case of a 2D rectangular sheet, a hollow cylinder and a hollow sphere of one coherence length thickness is calculated by the quantum mechanical electron-phonon scattering matrix, and a series of collective lattice vibrations in the surface state. We will show that being extremely thin in a flat rectangular shape is not enough to significantly enhance the $T_c$ through phonon softening. However, if a curvature is added, $T_c$ can be strongly enhanced. The increase in $T_c$ with respect to the bulk is greatest in a hollow sphere, intermediate in a hollow cylinder and weakest for the rectangular sheet, when systems of identical length scale are considered. In addition, we find that the edge effect of such a 2D sheet has a strong broadening effect on $T_c$ in addition to the effect of order parameter phase fluctuations.

## 2 Introduction

Superconductivity in reduced dimensionality in the form of nano-structured or nano-patterned materials, such as in ultra-thin films [1-4], nanowires [5-7] or nanowire arrays [8-10] has been a field of intense research in the past few decades. The superconducting properties in a 2D system are effectively modified by the instability of the Cooper pairs due to the finite size effect [11,12,13]. It has been found that an enhanced electron phonon coupling can arise from nano-structuring under special circumstances, either because of the surface effect, which is especially important in ultrathin materials [14,15], or for example in carbon-based materials by rolling 2D graphene sheets into the form of thin carbon nanotubes [16,17,18]. Phonon softening is assumed to be the key element to explain the effective increase of the onset temperature in superconductors of finite size [18]. However, in a recent report on ultrathin Pb films, a decrease of $T_c$ compared to bulk Pb was observed instead, even at a thickness of a few atomic layers only [19]. It is indeed often found that $T_c$ is rather suppressed, while the critical field is greatly increased in ultrathin superconducting films [11,19]. This contrasts with the recent observation that the superconducting transition temperature of hollow Pb microspheres [14] was increased from the bulk $T_c$ value of 7.2K to 11.0K (1.53 times higher). A similar increased onset $T_c$, 11.3K, was found in ultrathin quasi-1D Pb nanowire arrays (1.57 times higher) [15]. In this article, we examine the feasibility of increasing the superconducting temperature by nano-structuring of up to 1.6 times the bulk $T_c$ by tuning the curvature of the 2D superconducting materials, in perfect agreement with the experimental observations, and we model how the edges affect the sharpness of the superconducting phase transition in 2D thin sheets.

## 3 Theory

We use an array 800 × 800 grid points to model a 2D rectangular sheet. Each grid point corresponds to the coherence volume of an s-wave superconductor at $T = 0$. In 3D, the lattice vibration can be described by the displacement of atoms from their equilibrium position: $u_j(\mathbf{R}, t)$, where $j$ represents the $j$th atom in the unit cell of mass $M_j$, $\mathbf{R}$ the atom position, $t$ the time and $F_{j,j'}$ are the spring forces between pairs of atoms. In the harmonic approximation, the motion equations are linear homogeneous differential equations:

$$M_j \frac{d^2 u^j}{dt^2} + \sum_{j',R} F_{j,j'} u^{j'} = 0$$

With the plane wave solution: $u_j(\mathbf{R}, t) = A_j \exp[-i(\omega t - \mathbf{k} \cdot \mathbf{R})]$

In the following we assume that we have an elemental superconductor with one atom per unit cell and we consider nearest neighbors only: $M \frac{d^2 u}{dt^2} + Ku = 0$.

where $M$ is the ionic mass, $K$ is the an effective spring coupling between neighboring ions and $u$ is the displacement vector of the lattice vibration. We furthermore assume that the system is isotropic within the plane, and the spring couplings along $x$ and $y$ directions are equal. Each coherence volume encloses more than one unit cell, depending on the ratio between the coherence length and lattice constant. In other word, the spring couplings at each grid refer to the collective lattice vibrations within the coherence volume. We assign the collective spring constant to be 1 for simplicity and use a simple Einstein model with a mean collective vibrational frequency $\omega$, which is solvable if the mass per grid is given. This approach is justified by the fact that in many superconductors, such as e.g. MgB$_2$ [20] the phonons responsible for superconductivity are represented by a rather narrow group of phonon frequencies that can be reasonably well described by an Einstein model [21]. A classical approach is used to calculate the effective spring constant along the corresponding column and row at the grid $(i,j)$. A simplified 8 × 8 system in Figure 1 shows the blue interactive members with respect to the grid $(i, j)$. The reciprocal of the net spring coupling in each blue region equals to the sum of the reciprocal spring constants at each grid. Applying vector addition to all blue regions, we can find out the overall spring constant at the grid point $(i, j)$. A comparison of the net stiffness constant between $k(x,y)$ in the 2D rectangular film and $k(x,y,400)$ in the 3D rectangular block has been made carefully. Vector addition is used to compute the $R(x,y) = \dfrac{k(x,y,400)}{k(x,y)}$ before proceeding to calculate the electron phonon couplings and $T_c$ distribution. It is found that $R(x,y)$ is always larger than 1 because of the phonon softening.

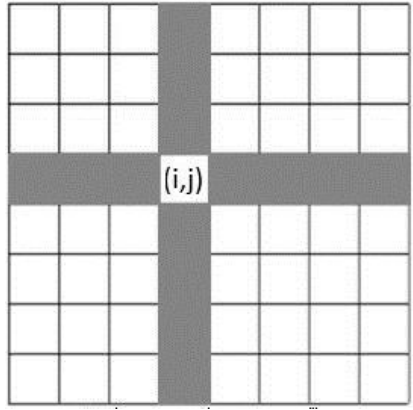

Figure 1: The grid members in blue color are considered as the interacting components with respect to the grid $(i,j)$. Apply the standard series and parallel classical mass – spring system to estimate the effective spring constant at the grid.

The obtained $\omega$ space aims at determining of the electron phonon scattering energy [17]

$$H_{e-ph} = \langle \psi_i | V | \psi_j \rangle$$

The electron wavefunction $\psi$ is interacting with the ionic Coulomb potential $V$. For small amplitude vibration ($\mathbf{u}_i = \mathbf{R}_i - \mathbf{R}_i^0$) and applying Bloch's theorem to the periodicity of the lattice position $\mathbf{R}_i^0$, the electron phonon interaction is calculated based on the conservation of momentum $\mathbf{q} = \mathbf{k} - \mathbf{k}' + \mathbf{G}$, where $\mathbf{G}$ is a reciprocal lattice vector, $\mathbf{k}$ and $\mathbf{k}'$ refer to electronic momentum states before and after scattering. [22] Applying 2nd quantization to the small displacement $\mathbf{u}$, the electron phonon coupling [23] is expressed as

$$H_{e-ph} = \sum_{kk'\sigma\lambda} g_{kk'\lambda} c_{k'\sigma}^{\dagger} c_{k\sigma} ((a_\lambda(\mathbf{q}) + a_\lambda^{\dagger}(-\mathbf{q}))$$

where [23,24], $a_{q\lambda}$ is defined as the phonon mode in which $a_{q\lambda}^{\dagger} a_{q\lambda} + \dfrac{1}{2}$ is the quantum number, $\lambda$ is the polarization and $c_{k\sigma}$ is the consequence of the expressing $\psi(r)$ to a linear combination of the eigenfunctions $\phi_k(\mathbf{r})$.

i.e. $\psi(r) = \sum_{k} c_{k\sigma} \phi_{k}(\mathbf{r})$,

$$g_{kk'\lambda} = (W_{kk'} \cdot e^{\lambda}(\mathbf{q}))\sqrt{\frac{N}{2M\omega(\mathbf{q})}}$$

$$W_{kk'} = \int d^{3}r \phi_{k'\sigma}^{*}(\mathbf{r})\phi_{k\sigma}(\mathbf{r})\nabla_{R_{i}^{0}}V(\mathbf{r}-\mathbf{R}_{i}^{0})$$

$N$ is defined as normal modes of vibration. e.g. If a three dimensional solid with 1 atom per unit cell is considered, there are $3N$ normal modes which presumably gives 3 acoustic branches. The polarization vector $e^{\lambda}(\mathbf{q})$ fulfills the following orthogonal requirements.

$$\sum_{\alpha} e_{\alpha}^{\lambda}(k) \cdot e_{\alpha}^{\lambda'}(k) = \delta_{\lambda\lambda'}$$

The Coulomb charge number $Z$ (a scalar) is extracted from the potential $V$ and implies that the electron phonon scattering increases with ionic charge. Any time a phonon of momentum $\mathbf{q}$ is excited, it generates charge density fluctuations arising from the positively charged ion. It will interact with the electron further, which requires us to account for the screening effect of the ionic charge. Therefore, the term, $g_{kk'\lambda}$, needs to be modified by the dielectric factor $\varepsilon$ in order to encounter for the screening effect in an isotropic system [23]

$$g_{eff}(\mathbf{q},\omega) = g(\mathbf{q})/\varepsilon(\mathbf{q},\omega)$$

Here $g(\mathbf{q})$ is the effective electron phonon coupling constant in an isotropic system. The electron phonon interaction in an isotropic system can be further organized as below

$$H_{e-ph} = \sum_{kk'\sigma\lambda} \frac{1}{\varepsilon(q,\omega)} \{[(\int d^{3}r \phi_{k'\sigma}^{*}(\mathbf{r})\phi_{k\sigma}(\mathbf{r})\nabla_{R_{i}^{0}}V(\mathbf{r}-\mathbf{R}_{i}^{0})) \cdot e^{\lambda}(\mathbf{q})]\sqrt{\frac{N}{2M}}\left(\frac{M}{K_{net}(q)}\right)^{0.25} c_{k'\sigma}^{\dagger}c_{k\sigma}((a_{\lambda}(\mathbf{q})+a_{\lambda}^{\dagger}(-\mathbf{q}))\}$$

Therefore $\dfrac{H_{e-ph}(x,y,400)}{H_{e-ph}(x,y)}$ is proportional to $(\dfrac{1}{R(x,y)})^{0.25}$.

The ratio of superconducting transition temperature between any grid is proportional to their electron phonon interaction ratio. Once the electron phonon coupling is known, the $T_c$ can be compared at any point on the surface. This relationship can be derived by using the pairing Hamiltonian.

The ground state of the BCS wavefunction $|\psi_{G}\rangle$ is [28]

$$|\psi_{G}\rangle = \prod_{k=k_{1}...k_{M}} \left(u_{k} + v_{k}c_{k\uparrow}^{*}c_{k\downarrow}^{*}\right)|\phi_{0}\rangle$$

where $|\phi_0\rangle$ is the vacuum state with no particles present, the probability of the pair $(k\uparrow,-k\downarrow)$ being occupied in state k is $|v_k|^2$, In other word, $|u_k|^2$ refers to the probability that is unoccupied since $|u_k|^2+|v_k|^2=1$. The creation operators, $c_{k\uparrow}^*$ and $c_{k\downarrow}^*$, correspond to spin up and down, respectively.

The pairing Hamiltonian [28], $H_p = \sum_{k\sigma} E_k n_{k\sigma} + \sum_{kl} V_{kl} c_{k\uparrow}^* c_{k\downarrow}^* c_{l\uparrow} c_{l\downarrow}$, is important to find out the connection between the superconducting energy gap and the electron phonon interaction, where $E_k$ is single particle energy relative to the Fermi energy, $n_{k\sigma}$ is particle number operator, $\sigma$ is spin index and the interaction term $V_{kl}$ scatters from a state with $(l\uparrow,-l\downarrow)$ to $(k\uparrow,-k\downarrow)$. The first term on the right is linked to kinetic energy (*KE*) and Fermi energy. Making use of the standard characteristic anticommutation relation of fermion operators [28], the mean number of particles $\bar{N}_n$ equals to $\sum_k 2|v_k|^2$ while $\langle(N_n-\bar{N}_n)^2\rangle = \sum_k 4u_k^2 v_k^2$ [28]. By setting $\delta\langle\psi_G|H_{pair}|\psi_G\rangle=0$ in combination to the mean number of particles, they yield

$$\langle KE - \mu N_n \rangle = 2\sum_k E_k |v_k|^2$$

$$H_{e-ph} = \langle V_{int}\rangle = \sum_{kl} V_{kl} u_k v_k^* u_l^* v_l$$

Following the variational method approach to take $u_k$ and $v_k$ to be real [28] we obtain $H_p = 2\sum_k E_k |v_k|^2 + \sum_{kl} V_{kl} u_k v_k u_l v_l$.

In the next step we compute $\frac{\partial}{\partial\theta_k}\langle\psi_G|H_{pair}|\psi_G\rangle=0$ in which we define $u_k=\sin\theta_k$ and $v_k=\cos\theta_k$ since $|u_k|^2+|v_k|^2=1$. Finally, the energy gap $\Delta_k$ is expressed as [28]

$$\Delta_k = -\sum_l V_{kl} u_l v_l$$

where $\Delta_k$ is essentially independent of $k$ and will be abbreviated as $\Delta$ [28].

Multiplying $-\sum_k u_k v_k$ to both sides of the above equation, i.e. $-\Delta\sum_k u_k v_k = \sum_{kl} V_{kl} u_k v_k u_l v_l$ the energy gap can be written as $\Delta = \dfrac{H_{e-ph}}{-\sum_k u_k v_k}$.

The energy gap $\Delta(0K)$ per particle equals to $1.76 k_B T_c$ [28] where $k_B$ is the Boltzmann constant. Assuming the Debye frequency to be independent of the grid position, the microscopic $T_c$ ratio between any pair of grid points can be obtained by the corresponding ratio in $H_{e-ph}$ at $T \sim 0K$, i.e. $T_{c(A)}/T_{c(B)} = \Delta_{(A)}/\Delta_{(B)} = H_{e-ph(A)}/H_{e-ph(B)}$ where ($A$) and ($B$) refer to any grid number.

The superconducting energy gap is a form of energy, and thus the $T_c$ is proportional to the gap. Taking the mean value of microscopic $T_c$ ratios on the grid points yields an overall macroscopic $T_c$ ratio, i.e. $\langle\Delta(0)\rangle = 1.76 k_B \langle T_c \rangle$. $\langle T_c \rangle$ refers here to the macroscopic $T_c$. As a result, by averaging the energy gaps the mean $T_c$ ratio can be derived according to the second law of thermodynamics.

### 3.1 Phonon softening in the rectangular superconducting film

Given that the thickness of the film equals to the coherence length, we are going to find out the $T_c$ distribution of the 2D rectangular superconducting film. The ratio of $H_{e-ph}$ is calculated with help of the $R(x, y)$ values and the $T_c$ ratio is obtained.

Table 1 shows the average $T_c$ of the 2D rectangular superconducting film in the different regions. The average $T_c$ ratio in the 800 x 800 film equals to 1.02 times the bulk $T_c$, because in 2D there are only 4 neighbors compared to the 6 in 3D. The $T_c$ is changing towards the edge, because of the spatial dependence of the electron phonon couplings. However, the $T_c$ is definitely increased by a ratio of 1.05 by phonon softening along the 4 edges. This is a consequence of the weakened spring constant or the slower vibrational frequencies at the edges. The highest $T_c$ occurs at the corners (ratio = 1.34) as shown in Table 1.

Table 1: The increase of $T_c$ due to phonon softening in the 800 x 800 superconducting film provided as the ratio $T_c/T_c^{bulk}$.

| Regions | The average $T_c$ ratio |
| --- | --- |
| Corners | 1.34 |
| Edges | 1.05 |
| The entire film | 1.02 |

### 3.2 Broadening effect of the superconducting phase transition due to the edge effect

Despite of the small area fraction with different $T_c$s that exist outside the central region, the effect of the finite size in presence of edges and corners is already sufficient to broaden the superconducting transition, as we will demonstrate in the following by deriving the specific heat anomaly at $T_c$ under the influence of the edge effect. The temperature dependence of the heat capacity $C_{es}$ of the 3D s-wave superconductors can be described by $C_{es}(T_c,T) = AT^{-1.5}e^{-3.52T_c/T}$. The broadening effect due to edge effect can be modelled by the summation of the heat capacity as a function of temperature at various $T_c$s in the 2D plane. Here $A$ refers to the total numbers of grids staying at one particular $T_c$.

Figure 2 shows the broadening effect on the specific heat anomaly due to edge effect. The sharpness is remarkably reduced in a 800 × 800 grid in the presence of a spatial variation in the electron phonon couplings. Note, that our approach is a mean field approach, and the strong 2D-XY phase fluctuations that are induced in 2D superconducting films associated with the Berezinskii-Kosterlitz-Thouless (BKT) transition [25] are neglected in our model. Nevertheless, the broadening from the edge effect is already significant. The ratio of the onset temperature in the 800 × 800 film is slightly increased to 1.06 due to the phonon softening.

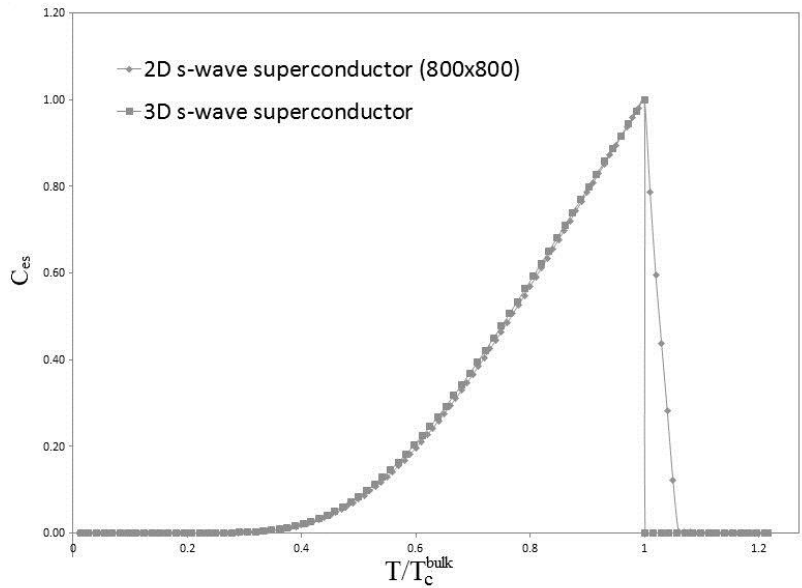

Figure 2: The normalized specific heat anomaly for a 2D 800 × 800 grid in comparison to 3D case. The superconducting transition is broadened by the variation of the electron phonon scattering energy in the 2D material of finite size.

### 3.3 The tighter bound Cooper pairs in the hollow sphere and cylinder

Following the same length scale, the superconducting transition temperature of a hollow sphere (Radius: 127, thickness: 1) is studied. The $T_c$s at every point on the surface are identical because of rotational symmetry and therefore the variation of $T_c$s observed in the case of the rectangular sheet disappears eventually. Each point refers to the coherence volume. The angle between the nearest neighbors in such a hollow sphere is 7.8 x 10$^{-3}$ in radian numerically. The electron phonon scattering on each point (e.g. the black dot as an example) along the surface in Figure 5, is calculated by the vector sum of the tangential components represented by the four black arrows, with help of the concept of the series of a mass – spring system. In the model, every grid point is treated as one collective mass, and therefore we utilize a simplified point mass or charge approach to treat the collective mass. It aims at modifying the Coulomb potential $U$ from the flat to spherical shape [17] with help of the corrected ionic charge number $Z_{corrected}$.

$$Z_{corrected} = Z\left(\frac{U_{sphere}}{U_{flat}}\right)$$

Here we assume the electron is moving around in a distance at $L/4$ along the radial axis, where $L$ is the separation between the nearest grid points. The potential energy in different curvatures is obtained by the vector sum of the potential energy between the electron and three nearest grid points.

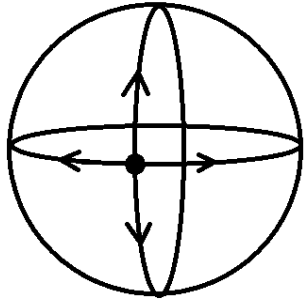

Figure 3: The superconducting transition temperature on each point on the surface is estimated by four tangential components of the electron phonon couplings (marked by the 4 black arrows). The resultant electron phonon coupling of each dot is influenced by the four series of vibrational points.

The electron phonon scattering energy is inversely proportional to the square root of the phonon vibrational frequency, while the vibrational frequency is proportional to the square root of the spring constant. With the calculated $Z_{corrected}$, we apply the same algorithm as in part A to obtain the average $T_c$ of the hollow sphere, the ratio is,

$$T_c^{sphere(2D)} / T_c^{cube(3D)} = 1.69$$

On the same length scale, the superconducting transition temperature of a hollow cylinder (Length: 800, Radius: 127, Thickness: 1) is investigated. After the $Z_{corrected} = Z\left(\frac{U_{cylinder}}{U_{flat}}\right)$ is obtained in the hollow cylindrical coordinate, we compute the electron phonon couplings along the $z$ axis and the curved $\theta$ component, respectively, in order to find out the enhancement of the superconducting transition temperature. Due to the symmetry along the angular component, the $T_c$ does not depend on $\theta$. It only varies along the $z$ axis and the average value is

$$T_c^{cylinder(2D)} / T_c^{cube(3D)} = 1.20.$$

This is a significant enhancement of $T_c$ induced by the curvature. Therefore, the effect of curvature in combination to the effect of phonon softening in the surface layer can be identified as the major reason of the observed $T_c$ enhancement that has been observed in Pb microspheres [14] and ultrathin Pb nanowire arrays [15].

## 4 Discussion

A stiffer spring constant causes a higher vibrational frequency. In Figure 1, each interior grid is attached to four series of the classical springs, where the minimum effective spring constant exists very close to the edge. Phonon softening occurs always at the edge, which causes a weaker spring constant. The weakening effect of the superconducting transition temperature comes from the insufficient scattering time between electrons and phonons. As a result, the electron phonon couplings in the central region are weaker relative to the edges and corners. The edge in Table 1 is associated with an obvious phonon softening resulting in a ratio $T_c/T_c^{bulk} = 1.05$. This is a consequence of the fact that the four series of springs are replaced by three. Only two series of springs link to the corner and it causes the much higher local $T_c$ at the corners ($T_c/T_c^{bulk} = 1.34$).

As a confirmation of recent experimental results [26] it is found that the superconducting phase transition of the trimmed film with its edges is sharper than in untrimmed films. A comparison is made between the bulk superconductor and the trimmed superconducting film in Figure 2. The $T_c$ distribution in the 2D superconducting material is much broader than in the 3D case. Although the broadness of the phase transition between the untrimmed (less sharp transition) and trimmed (sharper transition) superconductors [27] is much more obvious, the part B in this article discovers another source to influence the sharpness of the superconducting phase transition in low dimensional materials. The variation of the electron phonon couplings of the trimmed 2D superconducting film broadens the superconducting transition compared to bulk superconductors.

Being thin can only cause a minor increase of $T_c$ of around 1.05 times exceeding the bulk value [19], while the presence of a curvature [14,15] is able to increase it significantly up to a factor of $T_c/T_c^{bulk} = 1.6$. Due to the softer tangential spring components on the curved surface, the electron phonon couplings become larger. The $T_c$ ratio of $T_c/T_c^{bulk} = 1.69$ can be reached when the rectangular sheet (length 800 × 800, thickness 1) is reshaped to a hollow sphere (radius: 127, thickness: 1) on an identical length scale. It is consistent with the experimentally observed $T_c$ ratio [14] in thin hollow Pb microspheres ($T_c/T_c^{bulk} = 11.02/7.19 = 1.53$).

# 5 Conclusion

In this paper, a simple theoretical model is constructed to study the phonon softening in 2D superconductors of finite size. A spatial dependence of the electron phonon coupling near the edges is discovered to be another source to broaden the superconducting phase transition in the 2D plate in addition to the effect of strong phase fluctuations in reduced dimensionality. We could demonstrate with our model that the surface phonon softening in an ultrathin superconductor can be enhanced significantly upon introducing a curvature, in perfect agreement with recent observations of $T_c$ enhancements in ultrathin Pb nanowires [15] and hollow Pb microspheres [14].